# Weighted Contourlet Parametric (WCP) Feature Based Breast Tumor Classification from B-Mode Ultrasound Image

Shahriar Mahmud Kabir[a]*, Md. Sayed Tanveer[a], ASM Shihavuddin[a],

Mohammed Imamul Hassan Bhuiyan[b]

[a] Department of Electrical and Electronic Engineering, Green University of Bangladesh, Bangladesh
[b] Department of Electrical and Electronic Engineering, Bangladesh University of Engineering Technology, Bangladesh

**Abstract**

Automated detection of breast tumor in early stages using B-Mode Ultrasound image is crucial for preventing widespread breast cancer specially among women. This paper is primarily focusing on the classification of breast tumors through statistical modelling such as Rician inverse Gaussian (RiIG) pdf of contourlet transformed B-Mode image of breast tumors which is not reported yet in other earlier works. The suitability of RiIG distribution in modeling the contourlet coefficients is illustrated and compared with that of Nakagami distribution. The proposed method consists of pre-processing to remove the speckle noise, segmentation of the lesion region, contourlet transform on the B-Mode Ultrasound image and using the corresponding contourlet sub-band coefficients and the RiIG parameters, production of contourlet parametric (CP) images and weighted contourlet parametric (WCP) images. A number of geometrical, statistical, and texture features are calculated from B-Mode and the contourlet parametric images. In order to classify the features, seven different classifiers are employed. The proposed approach is applied to two different datasets (Mendeley Data and Dataset B) those are available publicly. It is shown that with parametric images, accuracies in the range of 94-97% are achieved for different classifiers. Specifically, with the support vector machine and k-nearest-neighbor classifier, very high accuracies of 97.2% and 97.55% can be obtained for the Mendeley Data and Dataset B, respectively, using the weighted contourlet parametric images.The reported classification performance is also compared with that of other works using the datasets employed in this paper. It is seen that the proposed approach using weighted contourlet parametric images can provide a superior performance.

---

*Corresponding author, (Shahriar Mahmud Kabir), Department of Electrical and Electronic Engineering, Green University of Bangladesh, 220/D, Begum Rokeya Sarani, Dhaka-1207, Bangladesh, Phone: +88-01764610728, e-mail address: shahriar.buet.msc@gmail.com

## 1. Introduction

Breast cancer in women is an important health problem for both developed and developing countries. A recent report by Cancer Statistics Center of American Cancer Society shows that among the estimated new cancer cases at 2020, cases of breast cancer number 1806590. It also shows that around 606520 cancer deaths are anticipated to ensue in the United States, of which breast cancer will number 279100, whereas lung and bronchus cancer will number 228820, prostate cancer will number 191930, colon cancer will number 104610, melanoma of the skin will number 100350 and urinary system cancer will number 159120. It is clearly shown that the number new cases from breast cancer is holding the top position in the year of 2020 [1].

Breast Ultrasound (US) imaging is one of the most promising tools to distinguish and classify breast tumors among the other imaging techniques such as Mammogram, MRI etc. Ultrasound is mainly a sound wave with frequencies higher than human audible frequencies (i.e. >20,000 Hz). Ultrasonic images are constructed by dispersing pulses of ultrasound into human tissue using a probe. In US imaging, the pulses echo off the body tissues having several reflection properties which are recorded and exhibited as an image. The B-Mode or brightness mode image then shows the acoustic impedance of a cross-section of tissue in two-dimensions.

A lot of research had been done and are still being done for the sake of differentiating malignant breast tumors from the benign ones. In 2002, Karla Horsch [2] used the depth-to-width ratio (DWR) of the region of lesion, the normalized radial gradient (NRG), auto correlation in depth of lesion region (COR) and minimum side difference (MSD) of the lesion boundary for the detection of breast tumors. In 2007, Wei-Chih Shen [3] presented a computer-aided diagnostic (CAD) system where few geometric features such as shape class, orientation class, margin class, lesion boundary class, echo pattern class and posterior acoustic feature class are used. They reported an accuracy is 91.7%. However, in their work, the segmentation of lesion was performed in both manually and automatically from the normal breast tissue, making it complicated for vast data of US images. In recent years, multi-resolution transform domain-based methods using US images showed higher promise in automatic breast Tumor classification task. In 2017, Sharmin R Ara [4] employed an empirical mode decomposition (EMD) method with discrete wavelet transform (DWT) followed by a wrapper algorithm to obtain a set of non-redundant features for classifying breast tumors and reported an accuracy is 98.01% on their dataset. Unfortunately, traditional DWT has limited directional information with the directions being only along with horizontal, vertical, and diagonal dimensions. Eltoukhy [5] presented a comparative study between two multi-resolution transform domain-based techniques, namely wavelet and curvelet, for breast tumor diagnosis in digital mammogram images. Contourlet transform, another multi-resolution transform domain-based technique [6], has been shown to be providing more directional information, with various directional decomposition levels increase along with the increase of the pyramidal decomposition levels. It is also a better descriptor of arbitrary shapes and contours. Contourlet-based mammography mass classification is reported in [7,8]. It is to be noted that there are many concerns about performing mammography, which utilizes low-energy x-ray radiation, for regular checkups. Moreover, in mammography, many women have to go through unnecessary breast biopsies due to lack of specificity. In benign cases this figure is about 65%– 85% of unnecessary breast biopsies. This unnecessary biopsy causing patients emotional and physical burdens by increasing the unexpected cost of mammographic screening which can be avoidable [9]. For that reason, researchers have recently been putting their efforts in relatively safer approaches like ultrasonography and elastography. In [10] contourlet transform is employed on Ultrasound shear-wave elastography (SWE) images where contourlet-based texture features were used with Fisher classifier for classification purpose and reported an accuracy of 92.5%. Contourlet transform is also employed in [11] on B-mode US, shear wave elastography (SWE) and contrast-enhanced US (CEUS) images and reported an accuracy of 67.57%, 81.08% and 75% respectively. Both DWT and curvelet transform are not capable of providing a variety of directions and also do not have good directional selectivity in two dimensions as compared to the contourlet transform. For that reason, this work also uses the contourlet transform method.

Many researchers, rather than trying to extract various features from the original B-mode images, had tried to use statistical modeling such as Gaussian or Nakagami models to create parametric models of the images [12,13] and have found satisfactory results. The primary inspiration behind these types of statisitical methods is to mathematically model the scattering of sound waves through the tissues, which can provide more insight into the system and thus, provide more accurate features. Moreover, statistical modeling can describe the false positive (FP) and false-negative (FN) more precisely than spatial domain visual ultrasound images. Ming-Chih Ho [14] used Nakagami modelling to address the detection of liver fibrosis in rats, which might be different from breast tumor classification, but it does provide some validation to the usefulness of parametric imaging. Rician Inverse Gaussian (RiIG) distribution, proposed in [15], has been shown to be highly suitable for modelling the statistics of the contourlet coefficients. As this work uses contourlet transform, RiIG distribution has been used instead of Nakagami distribution to model the contourlet coefficient images. From the experiments, it was shown that features extracted from the RiIG parametric images provide more accuracy for breast tumor classification than the features extracted from B-mode images. A different approach of using statistical modelling is introduced here as weighted contourlet parametric (WCP) images, where the parametric images are multiplied with the multi-resolution transform domain subbands such as contourlet decomposed subbands. In our work, weighted contourlet parametric (WCP) images are used to extract features, which are then used for the breast tumor classification. Firstly, the original B-mode images are subjected to contourlet transform up to 6 pyramidal and corresponding 32 directional decomposition levels. For each sub-band, RiIG parameters are estimated locally to construct contourlet parametric (CP) images. The product of the contourlet parametric (CP) images and the respective contourlet sub-bands give the WCP images. From the results, it was proven that the features extracted from the WCP images provide the highest accuracy compared to the original B-mode images, B-Mode parametric images, contourlet transformed images and CP images. It is to be noted that, this work is the first one to investigate the effectiveness of WCP images for breast tumor classification.

In our proposed method, a number of statistical, geometrical and texture features are extracted from the WCP images and then are subjected to various classifiers such as the support vector machine (SVM), k-nearest neighbours (KNN), fitted binary classification decision tree (BCT), fitted error-correcting output codes (ECOC) model, binary Gaussian kernel classification model (BGKC), linear classification models for two-class (binary) learning with high-dimensional (BLHD), fitted ensemble of learners for classification (ELC) etc. The performance of the prior classifiers is tested on two datasets of US images for breast tumor classification and compared with existing methods. Among the classifiers, the SVM gives the best accuracy, which is higher as compared with those of other algorithms. It is to be noted that it might be possible to achieve even greater accuracy using advanced state-of-the art methods such as Deep Convolutional Neural Network for the classification task. In these methods, however, the features that played a vital role in this classification task remain unknown. The algorithm-based extraction of well-established features used in various previous works from the WCP images provides a more reliable and validated proof of the effectiveness of RiIG based weighted contourlet parametric (WCP) images. The main contributions of this work are as follows,

• This paper uses Rician inverse Gaussian (RiIG) distribution for statistical modeling, which is a mixture of Rician distribution and the inverse Gaussian distribution and is reported as a more appropriate model for ultrasound image statistical modeling [15]. This paper is the first one to show the effectiveness of the RiIG over Nakagami distribution for breast tumor classification.

• Weighted Contourlet Parametric (WCP) image, which is the multiplication of RiIG based contourlet parametric (CP) images and the respective contourlet sub-bands, has been investigated for the first time in this work. It has been found to provide better accuracy in breast tumor classification.

• The proposed method, having a manual feature selection process, provides a more concrete proof of the usefulness of both contourlet transform and RiIG based statistical modelling for breast tumor classification with two different publicly available datasets consisting of 413 B-mode US images. It will, in the future, pave the way to the use of more sophisticated techniques such as deep learning for even greater accuracy.

## 2. Materials and Methods

### 2.1. *Datasets employed in this study*

The proposed method is established by considering 413 clinical cases including 210 Benign cases and 203 Malignant cases collected from two different databases. Among 413 clinical cases, 250 were considered from database-I (Mendeley Data) and 163 were considered from database-II (Dataset B). The database-I is contributed by Paulo Sergio Rodrigues, available at (https://data.mendeley.com/datasets/ wmy84gzngw/1) [16]. In this database, there are 250 US images where 100 are fibroadenoma (benign) and 150 are malignant cases. All the images are stored in *.bmp format. The database-II consists of 163 US images that are stored in *.png format, available at (http://www2.docm.mmu.ac.uk/STAFF/m.yap/ dataset.php) [17]. In this database the lesion regions (i.e. tumor contours) of 163 clinical cases were identified by a radiologist and stored as binary images in a separate folder; while the B-Mode US images were stored in another folder. The pathological findings of these 163 lesions were categorized into fibroadenoma (FA), invasive ductal carcinoma (IDC), ductal carcinoma in situ (DCIS), papilloma (PAP), unknown (UNK), lymph node (LN) and lymphoma (LP) etc. Among them 110 are benign and 53 are malignant cases. The details of the diagnosis of these two databases are summarized in Table 1.

Table 1: Patient data summary

| Database-I | | | |
|---|---|---|---|
| Tumor type | No. of patients | No. of lesions | Method of confirmation |
| Fibroadenoma (Benign) | 91 | 100 | Biopsy |
| Malignant | 142 | 150 | Biopsy |

| Database-II | | | |
|---|---|---|---|
| Tumor type | No. of patients | No. of lesions | Method of confirmation |
| Cyst (Benign) | 65 | 65 | Biopsy |
| Fibroadenoma (Benign) | 39 | 39 | Biopsy |
| Invasive Ductal Carcinoma (Malignant) | 40 | 40 | Biopsy |
| Ductal Carcinoma in Situ (Malignant) | 4 | 4 | Biopsy |
| Papilloma (Benign) | 3 | 3 | Biopsy |
| Lymph Node (Benign) | 3 | 3 | Biopsy |
| Lymphoma (Malignant) | 1 | 1 | Biopsy |
| Unknown (Malignant) | 8 | 8 | Biopsy |

Total = 413

## 2.2. Pre-processing

In database-I and database-II, the best frames for each clinical cases were previously chosen and stored by a radiologist in *.bmp and *.png formats respectively. In database-I, the benign and malignant cases were predefined in separate folders and in database-II, the benign and malignant cases were predefined in an xlsx file. Moreover, in database-II, the Tumor regions were also predefined in a separate folder.

### 2.2.1. Speckle Reduction

Speckle noise is an inherent property in coherent imaging that consists of medical ultrasound (US), synthetic aperture radar (SAR), and optical coherence. It reduces the resolution, contrast and sharpness of US images and thus make the diagnostic vital details complicated. Thus, the speckle noise reduction from medical US images is a vital task especially in pre-processing step such as tumor segmentation for various biomedical image processing tasks. So, the establishment of an effective method for reducing speckle noise, the knowledge of the statistics of the speckle noise is a pre-requisite. A number of varied statistical models have appeared in the literature for modelling the speckle noise such as the Rayleigh, Rician, Nakagami, K-Homodyne, Gamma, Weibull, normal, log-normal, Bessel k and Rician inverse Gaussian distributions [18,19,20,21,22]. Speckle reduction methods are mainly executed to eliminate the speckle noise without interfering and degrading the feature properties of the image. In this proposed method, the speckle reduction task is performed by applying "pretrained denoising deep neural network", denoted as 'DnCnn', which is the only currently available pretrained denoising network trained for grayscale images only [23]. The despeckling performance of the DnCnn is exhibited in Fig. 1.

### 2.2.2. Normalization

After performing speckle noise reduction, standardization was performed on each image to bring the pixel values to zero mean and unit variance. Then the pixel values were clipped to keep them within [-3,3]. These new clipped images were then translated and scaled to integer values within [0,255]. After performing this normalization process on the images, the ROI auto segmentation algorithms in MATLAB were able to perform the segmentation more accurately and effectively. Moreover, it also allowed statistical modeling (i.e.

Nakagami, RiIG) of the images to be more accurate, as the images now had a better data distribution.

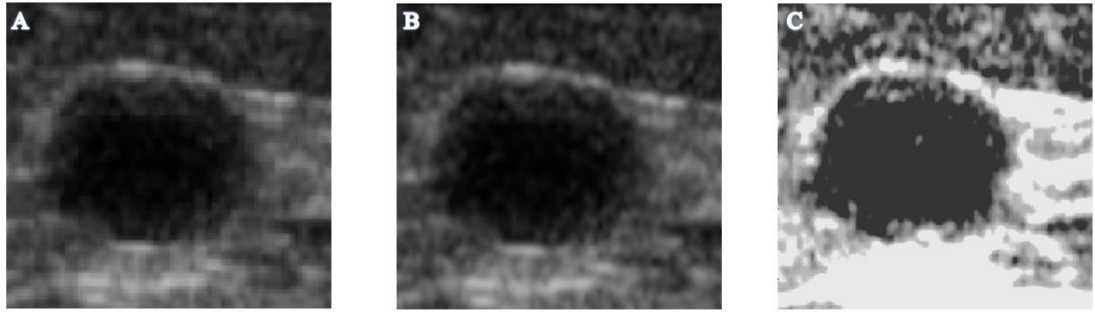

Fig. 1. Speckle reduction example of a benign mass (A) US B-Mode image (B) Denoised image by applying "pretrained denoising deep neural network" (C) Normalized Image.

2.2.3.    *Region of interest (ROI) Auto Segmentation*

The B-Mode images stored in database-I and database-II are in various sizes where the highest resolution of them is 224 x 224 pixels But, almost 50% of those images are contained pectoral muscle type huge noises in their background. Therefore, a shadow reduction operation is performed to minimize the unwanted portion of the image for the smooth detection of the region of interest (ROI), which is depicted in [24].

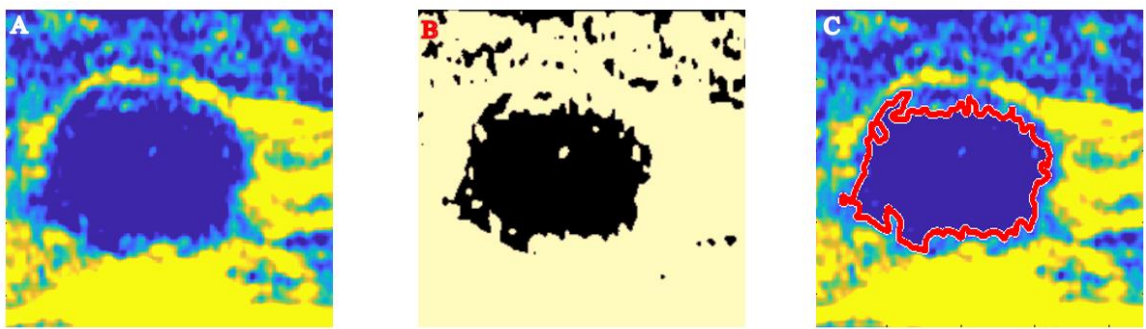

Fig. 2. Example of auto segmentation (A) Normalized B-Mode US image
(B) Corresponding binary image (C) Auto lesion boundary detection

It should be noted that, almost all the background information and lesion size must be preserved to ensure that no features will be suppressed with the background noise suppression. Next, the lesion boundary is fabricated using the Moore-Neighbor tracing algorithm modified by Jacob's stopping criteria [25]. This process requires a binary input image, specified as a 2-D logical or numeric matrix. The nonzero pixels of that binary image belong to an object and zero-valued pixels constitute the background as shown in Fig. 2.

### 2.2.4. *Contourlet Transform*

Traditional discrete wavelet transform (DWT) domain has limited directional information as only along with horizontal, vertical, and diagonal dimensions. On the other hand, the contourlet transform has a variety of arbitrary shapes and contours that are not limited to 3 dimensions. The contourlet transform is executed on the normalized B-Mode images which decouples the multiscale and the directional decompositions using a filter bank [6]. The conceptual theme of a contourlet transform is the decoupling operation that comprises a multiscale decomposition executed as pyramidal decomposition by a Laplacian pyramid and a following directional decomposition by engaging a directional filter bank. Fundamentally, the contourlet transform is constructed by the grouping of nearby wavelet coefficients, since they are locally correlated to ensure the smoothness of the contours. Therefore, a sparse expansion is obtained for natural images by first applying a multiscale transform, followed by a local directional transform to gather the nearby basis functions at the same scale into linear structures. Thus, it establishes a wavelet-like transform for edge detection and then a local directional transform for contour segment detection. The overall result is similar to an image expansion using basic elements that are more likely contour segments, and thus the name contourlets [21]. A performance comparison of DWT and contourlet transform in terms of better descriptor of contour segments are shown in Fig. 3. It is observed that, for DWT, the contour detection is performed with limited 3 dimensions and the detection become fader with the increase of decomposition levels. On the other hand, for contourlet transform, smooth contour detection with 32 dimensions is perceived, also detection become smoother with the increase of pyramidal decomposition levels.

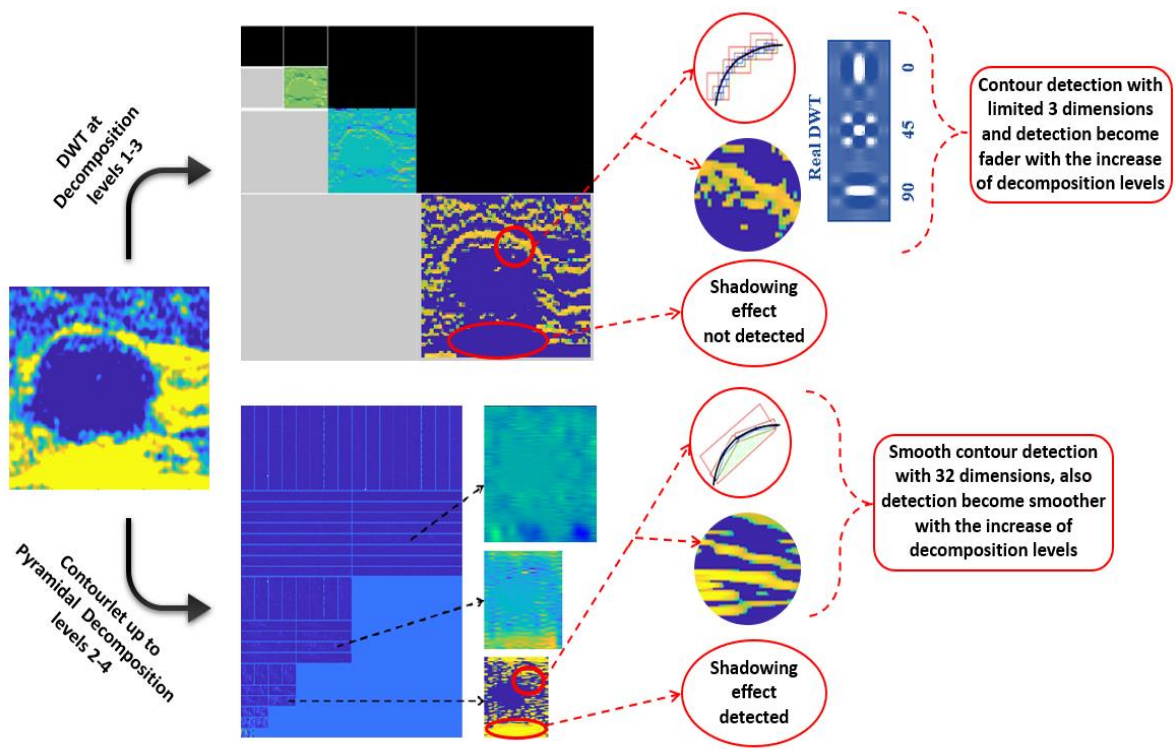

Fig. 3. DWT versus Contourlet scheme: illustrating the successive refinement of the two multi-resolution transform domains near a smooth contour

### 2.2.5. *Statistical Modeling/Contourlet Parametric (CP) Imaging*

*Rician Inverse Gaussian (RiIG) image*

The Rician Inverse Gaussian (RiIG) distribution is proposed by Torbjørn Eltoft [16]. It is a mixture of Rician distribution and Inverse Gaussian distribution. The pdf of Rician inverse Gaussian (RiIG) distribution is given by

$$P_{RiIG}(r) = \sqrt{\frac{2}{\pi}} \alpha^{\frac{3}{2}} \delta \, exp(\delta\gamma) \times \frac{r}{(\delta^2 + r^2)^{\frac{3}{4}}} K_{\frac{3}{2}}\left(\alpha\sqrt{\delta^2 + r^2}\right) I_0(\beta r) \qquad (1)$$

Where, α, β and δ are the three parameters of this *pdf*. α controls the steepness of the distribution, β regulates the skewness; β< 0 suggests skewed to the left, β> 0 suggests skewed to the right and δ is a dispersion parameter similar to the variance in the Gaussian distribution. A few realizations of the RiIG *pdfs* for a few selected values of the parameters are shown in Fig. 5, where the numbers in brackets in the upper right corner give the values of α, β, and δ for each curve (Fig. 4).

The contourlet parametric (CP) image is constructed from the RiIG parameter (δ) map, which is attained by employing a square sliding window to process the contourlet coefficient image. This process is depicted in [27], where the author used this process to construct Nakagami parametric image. It should be noted that in [26], the parametric images are obtained in the spatial domain, whereas we generated the images in the contourlet transform domain. The results observed in previous studies recommend that the most appropriate sliding window for constructing the parametric image is a square with a side length equal to three times the pulse length of the incident ultrasound [27].

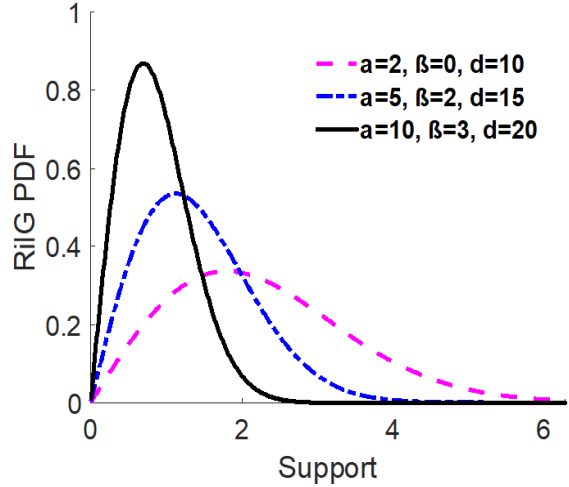

Fig. 4. Examples of the *pdf*s residing in the RiIG model. The numbers in the brackets correspond to α, β, δ respectively.

In this proposed method, the parametric imaging employed a 13x13 pixel sliding window within the contourlet sub-band coefficient image to analyze each local RiIG parameter ($\delta$). The employed sliding window size should be larger than the speckle and should discriminate variations of the local structure in tumors. The window was moved through the entire contourlet sub-band coefficient image in steps of 1 pixel, with the local RiIG parameter ($\delta$) assigned as the new pixel located at the center of the window at each position. This process yielded the RiIG parametric image as the map of RiIG parameter $\delta$ values. A comparison of Nakagami and RiIG parametric images is shown in Fig. 5, where the suitability of RiIG statistical model over Nakagami model is verified by percentile probability plot (*pp*-plot) [28,29].

2.2.1. *Statistical Modeling/Contourlet Parametric (CP) Imaging*

To reduce the computational time for constructing weighted contourlet parametric (WCP) images, we considered pyramidal decomposition levels 2, 3, and 4 in contourlet transform where those levels contain 8, 16, and 32 directional sub-bands respectively. Those are depicted as pyramidal level-2 directional level-4 (P2D4), pyramidal level-2 directional

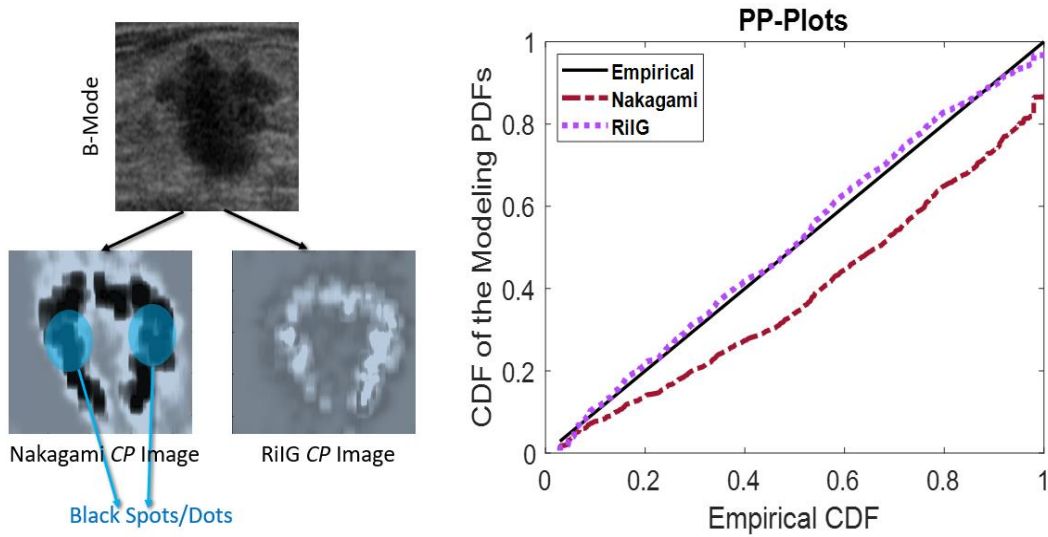

Fig. 5. Comparison of Nakagami and RiIG statistical modelling for image classification purpose (a) Nakagami and RiIG Contourlet Parametric (*CP*) images (b) Percentile probability plot (*pp-plot*) of Nakagami, RiIG and empirical Cumulative density functions (*cdf*s)

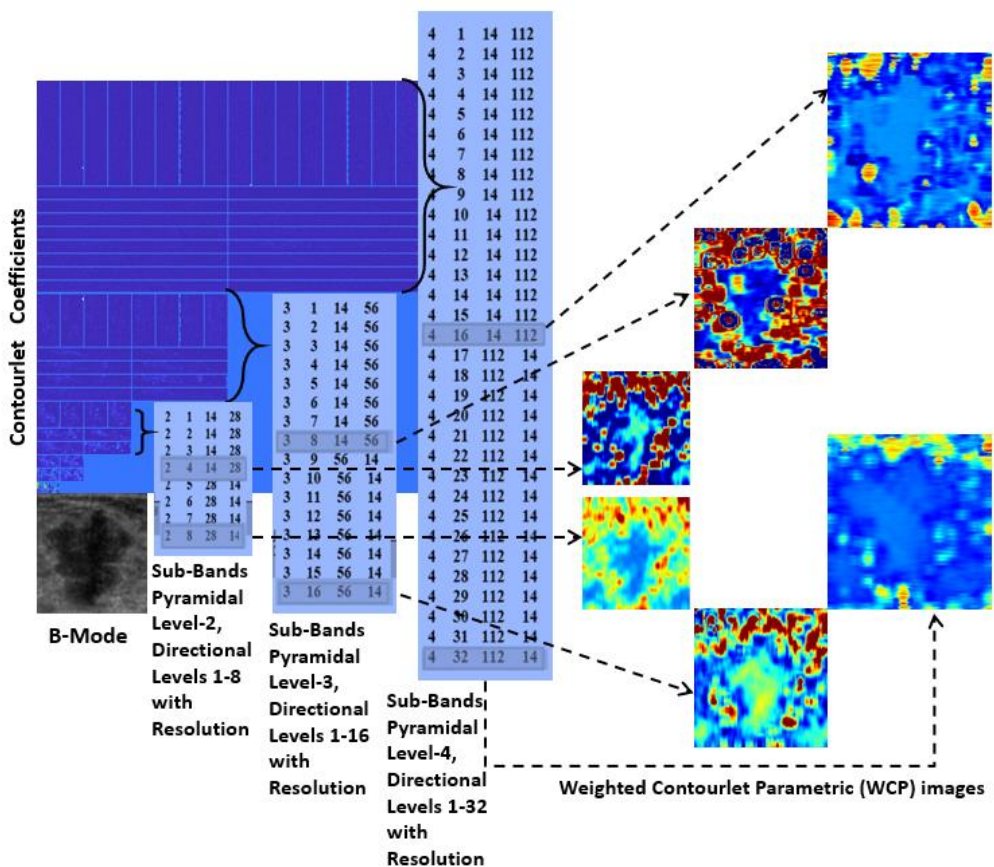

Fig. 6 Example of weighted contourlet parametric (WCP) images modeled by RiIG pdf at contourlet sub-bands pyramidal level-2 directional level-4 (P2D4), pyramidal level-2 directional level-8 (P2D8), pyramidal level-3 directional level-8 (P3D8), pyramidal level-3 directional level-16 (P3D16), pyramidal level-4 directional level-16 (P4D16) and pyramidal level-4 directional level-32 (P4D32).

level-8 (P2D8), pyramidal level-3 directional level-8 (P3D8), pyramidal level-3 directional level-16 (P3D16), pyramidal level-4 directional level-16 (P4D16) and pyramidal level-4 directional level-32 (P4D32) respectively. In this work, the directional sub-bands with the larger size in each level are considered, giving a total of 6-sub-bands from which the 6 RiIG parametric images are calculated. After obtaining the contourlet parametric (CP) images, each contourlet sub-band coefficient images are multiplied with their corresponding CP images to get weighted contourlet parametric (WCP) images. Features are extracted from these constructed WCP images to investigate the feasibility of tumor classification (Fig. 6).

### 2.3. Feature Extraction

A large set of ultrasound features does not inevitably guarantee the precise classification of breast Tumors. Sometimes it degrades the performance of the classifier. Moreover, most of the time it needed a high configuration system for computation. In this work, a small number of effective features are chosen based on the criterion used to assess the capability in both B-Mode and WCP images. The details of feature utilization are summarized in Table 2 and described in Appendix A.

### 2.4. Proposed Methodology

The proposed methodology can be summarized as shown in Fig. 7. At first, an ultrasound (US) image is considered for classification. In database-I & II the best frames are stored and identified as benign or malignant by a radiologist which is the benchmark of our proposed method. Therefore, it was not needed to choose a suitable frame. The two databases consisted of 413 B-Mode images. The images are at first de-speckled by applying "pretrained denoising deep neural network". Corresponding binary image is simulated from this filtered image (Fig. 2.b). The region of interest (ROI) (i.e. lesion region) is determined by the Moore-Neighbor tracing algorithm (Fig. 2.c) for feature extractions. The filtered and normalized B-Mode images are then subjected to contourlet transform in previously specified six sub-bands to reduce the computational time. These contourlet coefficients are then subjected to parametric imaging by statistical model RiIG pdf to obtain the contourlet parametric (CP) images. Total 2478 CP images are attained from 413 x 6=2478 contourlet coefficient images that were acquired from 413 B-Mode images. After that, these contourlet parametric (CP) images are multiplied with their corresponding contourlet coefficients to obtain weighted contourlet parametric (WCP) images. The region of interest (ROI) (i.e., lesion region) is determined for WCP images achieved results using Unitarian Rule that was to be as same as B-Mode images [13] for feature extractions. The total number of features from the data then reached 2478 x 20=49560. As using a large set of ultrasound features does not necessarily ensure improved quantitative classification of Breast Tumors and sometimes rather degrades the performance of a classifier; a filter or wrapper-based subset selection scheme is used to extract a set of non-redundant and more potential transform domain features. These reduced non-redundant features are used for validation and classification purposes using seven different classifiers as mentioned before. Finally, the classification decision, whether the tumor is benign or malignant, is decided by analyzing the maximum attribute of the seven classifiers. For this purpose, the total dataset is randomly divided into several groups such as training groups and testing groups with feature sets. 10-fold cross-validation process was applied for training and testing purpose, where every 10 clinical cases are considered as the test set for validation purpose, meanwhile remaining clinical cases are chosen as the training set for training purpose. The process is continued until the testing of each test set is accomplished from the cross-validation of all test sets. The performance of the proposed method is measured using the performance indices like accuracy, sensitivity, specificity, positive predictive value (PPV), and negative predictive value (NPV) etc. Later, the confusion matrices are obtained by measuring true positive (TP), true negative (TN), false positive (FP), and false-negative (FN) respectively where positive stands for malignant tumor and negative stands for a benign tumor.

Table 2: The feature utilization according to their capability in WCP images

| Feature | Method | Reference |
|---|---|---|
| Hypoechogenecity | $Hypo_{echo} = \dfrac{\sum_{i=1}^{N}\{pixel(i) < \overline{\|pixel(\iota)\|}\}\ ;\ N=pixels\ inside\ lesion}{\sum_{i=1}^{N}\{pixel(i) \geq \overline{\|pixel(\iota)\|}\}}$ | [30,31] |
| Microlobulation | $MicroLb_{echo} = \dfrac{\sum_{i=1}^{N}\{pixel(i) \geq \overline{\delta_{|pixel(\iota)|}}\}}{\sum_{i=1}^{N}\{pixel(i)\}}$ | |
| Homogeneous Echoes | $Homo_{echo} = \dfrac{\sum_{i=1}^{N}\{pixel(i) \geq 0\}}{\sum_{i=1}^{N}\{pixel(i)\}}$ | |
| Heterogeneous Echoes | $Hetero_{echo} = \dfrac{\sum_{i=1}^{N}\{pixel(i) < 0\}}{\sum_{i=1}^{N}\{pixel(i)\}}$ | |
| Taller Than Wide | $TW_{shape} = \left|\dfrac{Major\ axis\ length - Minor\ axis\ length}{100}\right|$ | |
| Microcalcification | $MicroCal_{echo} = \dfrac{\sum_{i=1}^{N}\{pixel(i) \geq \overline{\sigma_{|pixel(\iota)|}}\}}{\sum_{i=1}^{N}\{pixel(i) < \overline{\sigma_{|pixel(\iota)|}}\}}$ | [32,33] |
| Texture | $T_x = \left\{\dfrac{1}{N-1}\sum_{i=1}^{N}(x_i - \bar{x})^2\right\}^{\frac{1}{2}}$ | |
| Shape Class | $S_c = \dfrac{\sum_{p=1}^{N} D_{LE}(p)}{N}$ | [3] |
| Orientation Class | $O_c = \begin{cases} \theta & if\ 0 \leq \theta \leq \dfrac{\pi}{2} \\ \pi - \theta & if\ \dfrac{\pi}{2} < \theta \leq \pi \\ \theta - \pi & if\ \pi < \theta \leq \dfrac{3\pi}{2} \\ 2\pi - \theta & if\ \dfrac{3\pi}{2} < \theta \leq 2\pi \end{cases}$ | |
| Margin Class | $M_c = \dfrac{\sum Distance_{max}\{L(x,y)\}}{N_{Distance_{max}}\{L(x,y)\}} - \dfrac{\sum Distance_{min}\{L(x,y)\}}{N_{Distance_{min}}\{L(x,y)\}}$ | |
| Lesion Boundary Class | $L_c = \dfrac{\sum_{distance(L_{(x,y)outer})=1}^{k} I(L_{(x,y)outer})}{N_{L_{(x,y)outer}}} - \dfrac{\sum_{distance(L_{(x,y)inner})=1}^{k} I(L_{(x,y)inner})}{N_{L_{(x,y)inner}}}$ | |
| Echo Pattern Class | $E_{pc} = \dfrac{\sum_{L_{(x,y)}=1}^{N} G(L_{(x,y)})}{N_{L_{(x,y)}}}$ | |

| Feature | Method | Reference |
|---|---|---|
| Tilted Ellipse Radius | $R_{ellipse} = \dfrac{a+b}{2}$ | |
| Tilted Ellipse Perimeter | $P_{ellipse} \approx \pi\left\{3(a+b) - \sqrt{(3a+b)(a+3b)}\right\}$ | |
| Tilted Ellipse Area | $A_r = \pi a b$ | [34] |
| Tilted Ellipse Compactness | $CP_{ellipse} = \dfrac{P_{ellipse}{}^2}{A_{ellipse}} - 1.0$ | |
| PSD Peaks Vertical | *The frequencies where the first three peaks are detected obtained by power spectral density (PSD) estimation using all the pixel values on vertical sub-axis of the best fit tilted ellipse* | |
| Mean $klv$ | $klv = \max\limits_{x \in R} |F_e(x) - F_a(x)|$ | [28,29] |
| Percentile Probability Plot (*PP*-Plot) | $F_a(x)^t = \dfrac{2}{\pi} \arcsin\left(\sqrt{F_a(x)}\right)$  <br> $F_e(x)^t = \dfrac{2}{\pi} \arcsin\left(\sqrt{F_e(x)}\right)$ | [29] |
| Mean $klb$ | $KL(P_{emp}, P) = \int P_{emp}(x) \log_2 \dfrac{P_{emp}(x)}{P(x)} dx$ | [35] |

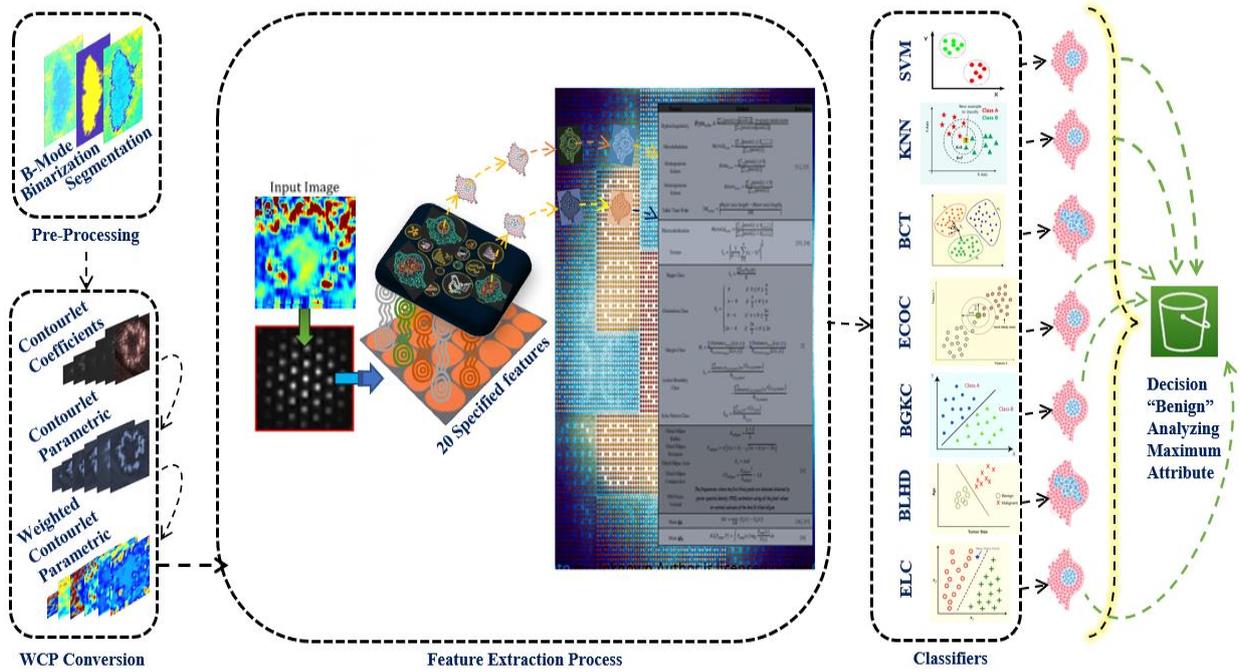

Fig. 7. The proposed architecture with step-by-step assessment

## 3. Experimental Results

In our proposed method, all the image processing tasks have been carried out using a workstation with an Intel 10th Generation core i7 processor with 6 cores, 2.4 GHz, 16 GB RAM using MATLAB R2020a (MathWorks Inc, USA) software. It typically took about 13 minutes to complete an epoch of training and predicting. The tabulated confusion matrices across the average of all 10-folds are presented in Table 4. It is observed that for the B-Mode image, the highest accuracy attained with the KNN classifier is about 92.02%.

Table 3: Classification performances for different types of images with databases-I and II

Accuracy (%) with database-I

| Classifier | B-Mode | Parametric | | Contourlet | Contourlet Parametric (CP) | | Weighted Contourlet Parametric (WCP) | |
|---|---|---|---|---|---|---|---|---|
| | | Nakagami | RiIG | | Nakagami | RiIG | Nakagami | RiIG |
| SVM | 92% | 81.6% | 93.2% | 92.4% | 84.8% | **94%** | 89.2% | **97.2%** |
| KNN | 91.6% | 80.8% | 92.4% | **92.8%** | 85.2% | 93.6% | 88.4% | 95.6% |
| BCT | 88.4% | 82.4% | 88.4% | 89.2% | 83.6% | 91.2% | 85.6% | 95.2% |
| ECOC | 90.4% | 80.4% | 90.8% | 91.6% | 85.2% | 92.4% | 86.8% | 94.8% |
| BGKC | 88% | 83.6% | 88.8% | 89.6% | 84.4% | 90% | 87.2% | 94% |
| BLHD | 89.2% | 82.8% | 90.4% | 90.4% | 84% | 90.8% | 87.6% | 94.4% |
| ELC | 92% | 82% | 92.8% | 92.4% | 84.8% | 93.2% | 89.6% | 96.8% |

Accuracy (%) with database-II

| Classifier | B-Mode | Parametric | | Contourlet | Contourlet Parametric (CP) | | Weighted Contourlet Parametric (WCP) | |
|---|---|---|---|---|---|---|---|---|
| | | Nakagami | RiIG | | Nakagami | RiIG | Nakagami | RiIG |
| SVM | 90.8% | 80.98% | **93.25%** | 91.41% | 85.89% | 93.87% | 87.73% | **97.55%** |
| KNN | **92.02%** | 82.82% | 91.41% | 92.64% | 87.12% | 93.25% | 88.96% | 96.32% |
| BCT | 88.34% | 84.05% | 88.34% | 89.57% | 83.44% | 90.8% | 86.5% | 95.09% |
| ECOC | 90.12% | 84.66% | 90.8% | 90.8% | 88.96% | 91.41% | 89.57% | 96.93% |
| BGKC | 87.73% | 80.98% | 88.96% | 88.96% | 84.66% | 90.57% | 87.12% | 94.48% |
| BLHD | 87.12% | 82.21% | 87.12% | 89.18% | 85.28% | 89.57% | 87.73% | 95.09% |
| ELC | 90.18% | 83.44% | 90.18% | 90.57% | 86.5% | 92.02% | 90.12% | 96.93% |

For B-Mode parametric image the highest attained accuracy is 93.2% with SVM classifier, for contourlet coefficient it is 92.8% with KNN classifier, for CP image it is 94% with SVM classifier and from WCP image the highest accuracy is attained with the SVM classifier, which is about 97.55% and is the highest. The performance of classification using different types of images are gathered in Table 3 and the confusion matrices for WCP images are gathered in Table 4.

Table 4: Classification performances of WCP images with databases-I and II

| Classifier | WCP Image analysis with database-I | WCP Image analysis with database-II |
|---|---|---|
| SVM | Predicted Class: Malignant TP=147, Benign FN=4 (Sensitivity=97.35%); Benign actual: FP=3, TN=96 (Specificity=96.97%); PPV=98%, NPV=96%, Accuracy=97.2% | Predicted Class: Malignant TP=53, Benign FN=3 (Sensitivity=94.64%); Benign actual: FP=1, TN=106 (Specificity=99.07%); PPV=98.15%, NPV=97.25%, Accuracy=97.55% |
| KNN | Malignant TP=146, FN=7 (Sensitivity=95.42%); FP=4, TN=93 (Specificity=95.88%); PPV=97.33%, NPV=93%, Accuracy=95.6% | Malignant TP=50, FN=2 (Sensitivity=96.15%); FP=4, TN=107 (Specificity=96.4%); PPV=92.59%, NPV=98.17%, Accuracy=96.32% |
| BCT | Malignant TP=143, FN=5 (Sensitivity=96.62%); FP=7, TN=95 (Specificity=93.14%); PPV=95.33%, NPV=95%, Accuracy=95.2% | Malignant TP=51, FN=5 (Sensitivity=91.07%); FP=3, TN=104 (Specificity=97.2%); PPV=94.44%, NPV=95.41%, Accuracy=95.09% |
| ECOC | Malignant TP=146, FN=9 (Sensitivity=94.19%); FP=4, TN=91 (Specificity=95.79%); PPV=97.33%, NPV=91%, Accuracy=94.8% | Malignant TP=53, FN=4 (Sensitivity=92.98%); FP=1, TN=105 (Specificity=99.06%); PPV=98.15%, NPV=96.33%, Accuracy=96.93% |
| BGKC | Malignant TP=144, FN=9 (Sensitivity=94.12%); FP=6, TN=91 (Specificity=93.81%); PPV=96%, NPV=91%, Accuracy=94% | Malignant TP=48, FN=3 (Sensitivity=94.12%); FP=6, TN=106 (Specificity=94.64%); PPV=88.89%, NPV=97.25%, Accuracy=94.48% |

| Classifier | WCP Image analysis with database-I | WCP Image analysis with database-II |
|---|---|---|
| BLHD | Predicted Class: Malignant / Benign; Actual Malignant: TP = 143, FN = 7, Sensitivity = 95.33%; Actual Benign: FP = 7, TN = 93, Specificity = 93%; PPV = 95.33%, NPV = 93%, Accuracy = 94.4% | Predicted Class: Malignant / Benign; Actual Malignant: TP = 50, FN = 4, Sensitivity = 92.59%; Actual Benign: FP = 4, TN = 105, Specificity = 96.33%; PPV = 92.59%, NPV = 96.33%, Accuracy = 95.09% |
| ELC | Predicted Class: Malignant / Benign; Actual Malignant: TP = 148, FN = 6, Sensitivity = 96.1%; Actual Benign: FP = 2, TN = 94, Specificity = 97.92%; PPV = 98.67%, NPV = 94%, Accuracy = 96.8% | Predicted Class: Malignant / Benign; Actual Malignant: TP = 51, FN = 2, Sensitivity = 96.23%; Actual Benign: FP = 3, TN = 107, Specificity = 97.27%; PPV = 94.44%, NPV = 98.17%, Accuracy = 96.93% |

In comparison, Pedro Acevedo-Contla [36] used a gray level concurrency matrix (GLCM) algorithm with lineal SVM to classify benign and malignant tumors using the dataset-I, yielding an accuracy of 94%. In another recent work [37] using also the same dataset-I, 94.8% accuracy is reported. So, the accuracy level obtained by the proposed method using the same database-I about 97.2% is satisfactory. S.Y. Shin [38], illustrated a neural network with faster R-CNN and ResNet-101 using database-II with added other databases. The author reported an accuracy of 84.5% using the same database-II. Our attained accuracy of 97.55% is significantly better. Michal Byra [39], presented an approach by the US to RGB conversion and fine-tuning using back-propagation. The author used database-II added with other databases and reported a specificity of 90% where our proposed method using database-II the specificity obtained 96.23% which is significantly better. Xiaofeng Qi [40], illustrated a novel approach of Deep CNN with multi-scale kernels and skip connections using database-II added with other datasets and reported an accuracy of 94.48%. In our proposed method using database-II, the accuracy achieved 97.55% which is significantly higher. The comparisons are summarized in Table 5.

Table 5: A comparison of selected studies in breast tumor classification task using databases-I and II

| Author (Year) | Novelty of Paper | Database | Classifier | Performance (accuracy in %) |
|---|---|---|---|---|
| Pedro Acevedo-Contla (2019), [36] | Gray level concurrency matrix (GLCM) algorithm | "Breast Ultrasound Image," Mendeley Data, vol.1, ver.1, 2017, DOI:10.17632/wmy84gzngw.1. Available: https://data.mendeley.com/datasets/wmy84gzngw/1 | SVM | ACC: 94% |
| Dennis Hou (2020), [37] | Portable device-based CNN architecture | "Breast Ultrasound Image," Mendeley Data, vol.1, ver.1, 2017, DOI:10.17632/wmy84gzngw.1. Available: https://data.mendeley.com/datasets/wmy84gzngw/1 | CNN | ACC: 94.8% |
| S. Y. Shin (2019) [38] | Neural Network with Faster R-CNN and ResNet-101 | Breast Ultrasound Lesions Dataset (Dataset B) Available: http://www2.docm.mmu.ac.uk/STAFF/m.yap/dataset.php + Other Database | R-CNN | ACC: 84.5% |
| Michal Byra (2019) [39] | US to RGB Conversion and fine-tuning using back-propagation | Breast Ultrasound Lesions Dataset (Dataset B) Available: http://www2.docm.mmu.ac.uk/STAFF/m.yap/dataset.php + Other Database | Deep CNN | SEN: 85% SPEC: 90% |
| Xiaofeng Qi (2019) [40] | Deep CNN with multi-scale kernels and skip connections. | Breast Ultrasound Lesions Dataset (Dataset B) Available: http://www2.docm.mmu.ac.uk/STAFF/m.yap/dataset.php + Other Databases | Deep CNN | ACC: 94.48% SEN: 95.65% |
| **Proposed Method** | RiIG based Weighted Contourlet Parametric (WCP) Image | "Breast Ultrasound Image," **Mendeley Data**, vol.1, ver.1, 2017, DOI:10.17632/wmy84gzngw.1. Available: https://data.mendeley.com/datasets/wmy84gzngw/1 Breast Ultrasound Lesions Dataset (**Dataset B**) Available: http://www2.docm.mmu.ac.uk/STAFF/m.yap/dataset.php + Other Databases | SVM, KNN, BCT, ECOC, BGKC, BLHD, ELC | **Mendeley Data**: ACC: 97.2% SEN: 97.35% SPEC: 96.97% **Dataset B**: ACC: 97.55% SEN: 94.64% SPEC: 99.07% |

## 4. Conclusions

In this paper, a new approach employing parametric images obtained from the contourlet transforms of ultrasound images of breast tumors. The Rician Inverse Gaussian (RiIG) distribution has been presented as an appropriate distribution for modeling the contourlet transform coefficients of B-mode ultrasound images of breast tumors. Contourlet transform is chosen due to its variety of directionalities, which is not limited to 3 dimensions like DWT, rather directional decomposition levels increase along with the increase of the pyramidal decomposition levels. The RiIG parameters obtained from the contourlet coefficients have been used to generate the contourlet parametric (CP) images. A number statistical, geometrical and texture features have been extracted from the non-parametric as well as weighted contourlet parametric (WCP) images. The WCP images are obtained by multiplying a contourlet parametric (CP) image with the corresponding contourlet coefficients. The performance of the proposed approach has been studied using two publicly available datasets and compared with the existing works. It has been shown that the proposed method can deliver a superior performance in accuracy with a high degree of accuracy, specificity and sensitivity. In future works, other multi-resolution transform domains such as Dual-Tree Complex Wavelet Transform (DTCWT) or Quaternion Wavelet Transform (QWT) and Curvelet transform will be employed and the experimentation will be carried out on a larger number of US B-Mode images. Moreover, investigations using the Deep Convolutional Neural Networks will be carried out.

# Appendix A

*Hypoechogenecity ($Hypo_{echo}$)*

The surrounding tissue of an area seems darker on ultrasound is known as hypoechoic. The hypoechoic tissue looks brighter/lighter shades of grey. When a mass is solid, rather than liquid then it means a hypoechoic mass. The hypoechoic lesion is not a cyst. Ultrasound utilizes sound-waves that has a property of echoic bounce back. The strong echoic bounce back appears brighter to the scanning probe on ultrasound, where the region absorbs sound wave don't have any echoic bounce back to the scanning probe, appears darker on the ultrasound image. Hyper means a huge echo those made something seems brighter, and hypo-echoic means the nodule that made something seems darker on the ultrasound [30,31]. A malignant breast mass includes a marked hypoechogenicity. By analyzing a WCP image, the hypoechogenicity can be measured as,

$$Hypo_{echo} = \frac{Total\ number\ of\ pixels\ having\ values < mean(normalization(pixel\ values\ inside\ lesion));\ N = pixels\ inside\ lesion}{Total\ number\ of\ pixels\ having\ values \geq mean(normalization(pixel\ values\ inside\ lesion))}$$

$$Hypo_{echo} = \frac{\sum_{i=1}^{N}\{pixel(i) < \overline{\|pixel(i)\|}\}\ ;\ N = pixels\ inside\ lesion}{\sum_{i=1}^{N}\{pixel(i) \geq \overline{\|pixel(i)\|}\}} \quad (2)$$

*Microlobulation ($MicroLb_{echo}$)*

The availability of very small tissues (1mm to 2 mm) is denoted as the lobulations on the surface. For a solid breast nodule those are considered as microlobulation. As the number of these microlobulations increase, the probability of malignancy also increases [30,31]. It is observed that in a simulated WCP image, the probably microlobulation can be measured as,

$$MicroLb_{echo} = \frac{Total\ number\ of\ pixels\ having\ values \geq mean(std(pixel\ values\ inside\ lesion))}{Total\ number\ of\ pixels\ inside\ the\ lesion\ region}$$

$$MicroLb_{echo} = \frac{\sum_{i=1}^{N}\{pixel(i) \geq \overline{\delta_{|pixel(i)|}}\}}{\sum_{i=1}^{N}\{pixel(i)\}} \quad (3)$$

*Homogeneous Echoes ($Homo_{echo}$)*

Since, the region on ultrasound that absorbs sound wave, don't have any echoic bounce back to the scanning probe, appears darker on the ultrasound image [31]. In this context, for WCP image the homogeneous echoes can be considered as the rate of the positive pixel values inside the lesion region as given as,

$$Homo_{echo} = \frac{Total\ number\ of\ pixels\ having\ positive\ values\ inside\ the\ lesion\ region}{Total\ number\ of\ pixels\ inside\ the\ lesion\ region}$$

$$Homo_{echo} = \frac{\sum_{i=1}^{N}\{pixel(i) \geq 0\}}{\sum_{i=1}^{N}\{pixel(i)\}} \quad (4)$$

*Heterogeneous Echoes ($Hetero_{echo}$)*

As the strong echoic bounce back appears brighter to the scanning probe on ultrasound images [31]. So, in the proposed method, the heterogeneous echoes for WCP images are considered as the rate of the negative pixel values inside the lesion region as given as,

$$Hetero_{echo} = \frac{Total\ Number\ of\ pixel\ having\ negative\ values\ inside\ the\ lesion\ region}{Total\ Number\ of\ Pixels\ inside\ the\ Lesion\ Region}$$

$$Hetero_{echo} = \frac{\sum_{i=1}^{N}\{pixel(i) < 0\}}{\sum_{i=1}^{N}\{pixel(i)\}} \quad (5)$$

*Taller Than Wide ($TW_{shape}$)*

This feature can play a vital role whenever a solid breast nodule is significantly show up taller-than-wide in geometric shape, then this increases the probability of malignancy. This 'tallness' geometric shape can be compared in sagittal or transverse dimensions (depth and width) which indicates that this may be a malignancy 'aggressive enough' to overcome normal breast tissue barriers and planes and grow vertically [30,31]. In this paper, this feature for WCP image is measured as,

$$TW_{shape} = \left|\frac{Major\ axis\ length - Minor\ axis\ length}{100}\right| \quad (6)$$

Here, the major axis means the Y-axis and the minor axis means the X-axis of the best-fit tilted ellipse in the lesion region (Fig. 8).

*Microcalcification ($MicroCal_{echo}$)*

The presence of small calcium deposits in the soft tissue of breast US image is known as breast microcalcifications. Breast microcalcifications often indicates the benign (non-cancerous) breast condition. Breast microcalcifications appear as white dots on ultrasound. This condition may appear as a result of genetic mutations somewhere in the breast tissue [32,33]. Analyzing WCP images, this feature is measured as,

$$MicroCal_{echo} = \frac{Total\ Number\ of\ Pixels\ having\ values \geq mean(variance(pixel\ values\ inside\ lesion))}{Total\ Number\ of\ Pixels\ having\ values < mean(variance(pixel\ values\ inside\ lesion))}$$

$$MicroCal_{echo} = \frac{\sum_{i=1}^{N}\{pixel(i) \geq \overline{\sigma_{|pixel(i)|}}\}}{\sum_{i=1}^{N}\{pixel(i) < \overline{\sigma_{|pixel(i)|}}\}} \quad (7)$$

*Texture ($T_x$)*

The combination of microcalcification textures with particular cell presentations gives a clearer picture of probable cancer malignancy [33]. It is observed that in a simulated multiplied image, the probable microcalcification texture can be measured as,

$$T_x = \left(\frac{1}{N-1}\sum_{i=1}^{N}(x_i - \bar{x})^2\right)^{\frac{1}{2}} \quad (8)$$

Here, N is the total number of pixels inside the lesion region, $x_i$ is the pixel values where $i = 1 \rightarrow N$ and $\bar{x}$ is the average of pixel values inside the lesion region.

*Shape Class ($S_c$)*

In this paper, the best-fit ellipse is used to roughly describe the lesion shape and considered as a baseline for measuring the degree of irregular shape [41] which is shown in Fig. 8. If *L(x1, y1)* is any pixel on the lesion boundary and *E(x2, y2)* is a crossing pixel on the best-fit tilted ellipse, measured by a ray that starts from the center of the best-fit tilted ellipse. The

variation between L(x1, y1) and E(x2, y2) is then estimated by their distance [3]

$$D_{LE} = \sqrt{(x_1 - x_2)^2 + (y_1 - y_2)^2} \tag{9}$$

Finally, the shape feature $S_c$ is measured as

$$S_c = \frac{\sum_{p=1}^{N} D_{LE}(p)}{N} \tag{10}$$

Here, $N$ is the total number of pixels on the lesion boundary, and $p$ is one of the lesion boundary pixels where, $p = 1 \rightarrow N$

*Orientation Class ($O_c$)*

This feature is measured by considering the angle of the major axis of best-fit tilted ellipse as shown in Fig. 9. The range parallel degree of lesion orientation should be measured between 0 and $\frac{\pi}{2}$. Thus, the orientation feature $O_c$ is estimated by fine-tuning the acquired $\theta$ into the range 0 and $\frac{\pi}{2}$ as depicted in [3],

$$O_c = \begin{cases} \theta & if\ 0 \leq \theta \leq \frac{\pi}{2} \\ \pi - \theta & if\ \frac{\pi}{2} < \theta \leq \pi \\ \theta - \pi & if\ \pi < \theta \leq \frac{3\pi}{2} \\ 2\pi - \theta & if\ \frac{3\pi}{2} < \theta \leq 2\pi \end{cases} \tag{11}$$

*Margin Class ($M_c$)*

The margin class is measured by the distance map [42]. This feature is significant for capturing the undulation and angular characteristics of the lesion boundary margin. For any pixel P(x, y) in the ROI, its eight distance neighbors are shown in Fig. 10, where '$i$' is limited from 1 to 13 pixels. It means the search limit is fixed up-to 13 pixels, from any pixel of lesion boundary to the nearest any pixel of best-fit tilted ellipse boundary.

$$Neighbor8\{L(x,y) \rightarrow E(x,y)\} = \{(x-i, y-i), (x, y-i), (x+i, y-i), (x-i, y), (x+i, y), (x-i, y+i), (x, y+i),$$
$$(x+i, y+i)\} \tag{12}$$

Here, $L(x, y)$ is any pixel on the lesion boundary and $E(x, y)$ is any pixel on the best-fit tilted ellipse boundary, where $i = 1 \rightarrow 13$. In this research by analyzing $i = 3, 6, 9$ and $13$; the best performance is attained considering 13 as the distance from the lesion to the ellipse boundary which is also validated in [3]

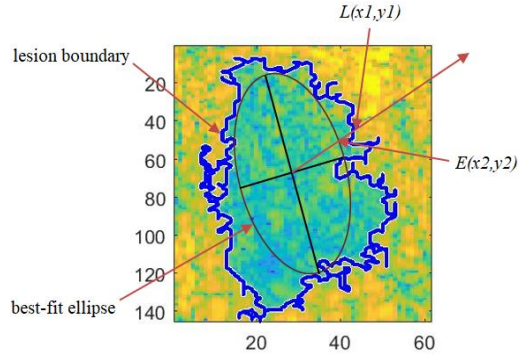

Fig. 8. The lesion shape class is estimated by the difference between the lesion boundary and the corresponding best-fit tilted ellipse.

$$Distance_{min}\{L(x,y)\} = Min[Distance\{Neighbor8(L(x,y) \rightarrow E(x,y))\}] + 1 \quad (13)$$

$$Distance_{max}\{L(x,y)\} = Max[Distance\{Neighbor8(L(x,y) \rightarrow E(x,y))\}] \quad (14)$$

$$M_c = \frac{\sum Distance_{max}\{L(x,y)\}}{N_{Distance_{max}}\{L(x,y)\}} - \frac{\sum Distance_{min}\{L(x,y)\}}{N_{Distance_{min}}\{L(x,y)\}} \quad (15)$$

Here, $N_{Distance_{max}}\{L(x,y)\}$ is the total number of maximum distances found for all pixels on lesion boundary and $N_{Distance_{min}}\{L(x,y)\}$ is the total number of minimum distances found for all pixels on the lesion boundary.

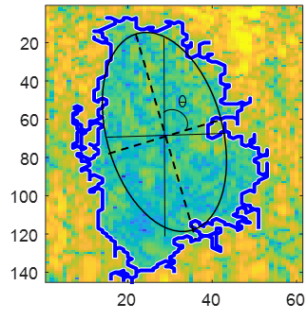

Fig. 9. The orientation class of a lesion is estimated by the angular characteristics of the major axis of best-fit tilted ellipse.

*Lesion Boundary Class ($L_c$)*

This feature is measured by estimating the degree of sudden interface across the lesion region boundary. The distance map shown in Fig. 11 is used to barrier the surrounding tissue and outer mass. The involving pixels inside the barrier of the surrounding tissue and outer mass, are considered at a distance less than or equal to 13 from the lesion boundary. The prior 13 is not a magic figure, rather it is observed and chosen as the best performer between 6 and 13 for WCP images which is also validated in [3] for B-Mode images. The average of the gray intensities of these two areas (i.e. from lesion boundary to outer mass ($\bar{I}_{outer}$) and from lesion boundary to surrounding tissue ($\bar{I}_{inner}$) inside the lesion region) with width k are defined in [3]

$$\bar{I}_{outer} = \frac{\sum_{distance(L_{(x,y)outer})=1}^{k} I(L_{(x,y)outer})}{N_{L_{(x,y)outer}}} \tag{16}$$

$$\bar{I}_{inner} = \frac{\sum_{distance(L_{(x,y)inner})=1}^{k} I(L_{(x,y)inner})}{N_{L_{(x,y)inner}}} \tag{17}$$

$$L_c = \bar{I}_{outer} - \bar{I}_{inner} \tag{18}$$

Here, the width k is considered as 6 by experiencing on WCP image in this study.

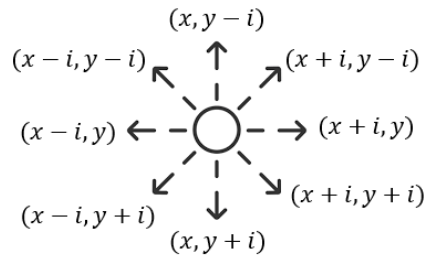

Fig. 10. The distance map of the lesion

*Echo Pattern Class ($E_{pc}$)*

The echo pattern class of a lesion is calculated by measuring the average gray intensity, $\bar{I}(L_{(x,y)})$ and the dissimilarity of lesion mass to represent the echo pattern features, respectively. The dissimilarity on pixel $L_{(x,y)}$ is measured by the gradient magnitude.

According to the definition of neighbors denoted by equation (12) and projected by Fig. 10, the Sobel gradients [42] on x-direction and y-direction of pixel $L_{(x,y)}$ are respectively defined as [3]

$$\bar{I}(L_{(x,y)}) = \frac{\sum_{L_{(x,y)}=1}^{N} I(L_{(x,y)})}{N_{L_{(x,y)}}} \quad (19)$$

$$G_X(L_{(x,y)}) = \{I(x-1, y-1) + 2*I(x-1, y) + I(x-1, y+1) - I(x+1, y-1) - 2*I(x+1, y) - I(x+1, y+1)\} \quad (20)$$

$$G_Y(L_{(x,y)}) = \{I(x-1, y-1) + 2*I(x, y-1) + I(x+1, y-1) - I(x-1, y+1) - 2*I(x, y+1) - I(x+1, y+1)\} \quad (21)$$

Then, the gradient magnitude on $L_{(x,y)}$ is defined as

$$G(L_{(x,y)}) = \sqrt{G_X(L_{(x,y)})^2 + G_Y(L_{(x,y)})^2} \quad (22)$$

Finally, the internal echo pattern class $E_{pc}$ is estimated by the average dissimilarity as

$$E_{pc} = \frac{\sum_{L_{(x,y)}=1}^{N} G(L_{(x,y)})}{N_{L_{(x,y)}}} \quad (23)$$

Here, $N_{L_{(x,y)}}$ denotes the number of pixels in the lesion region.

*Tilted Ellipse Radius ($R_{ellipse}$)*

The best-fit ellipse is simulated using the ellipse formula as

$$\left(\frac{X-X0}{a}\right)^2 + \left(\frac{Y-Y0}{b}\right)^2 = 1 \quad (24)$$

Here, $(X, Y)$ are the coordinates of the best fit tilted ellipse inside lesion boundary region, where $(X_0, Y_0)$ are the center of the best fit tilted ellipse; 'a' and 'b' are the sub axis radius of horizontal $(X)$ and vertical $(Y)$ sub axis respectively (e.g. Fig. 12). Then the radius is measured as,

$$R_{ellipse} = \frac{a+b}{2} \quad (25)$$

## Tilted Ellipse Perimeter ($P_{ellipse}$)

The accurate approximations to the perimeter of an ellipse is estimated as depicted in [34]. Accordingly, 'a' is considered as the length of the horizontal sub axis and 'b' is considered as the length of the vertical sub axis inside the best fit tilted ellipse (Fig. 12). So, this feature is calculated by,

$$P_{ellipse} \approx \pi \left\{ 3(a+b) - \sqrt{(3a+b)(a+3b)} \right\} \tag{26}$$

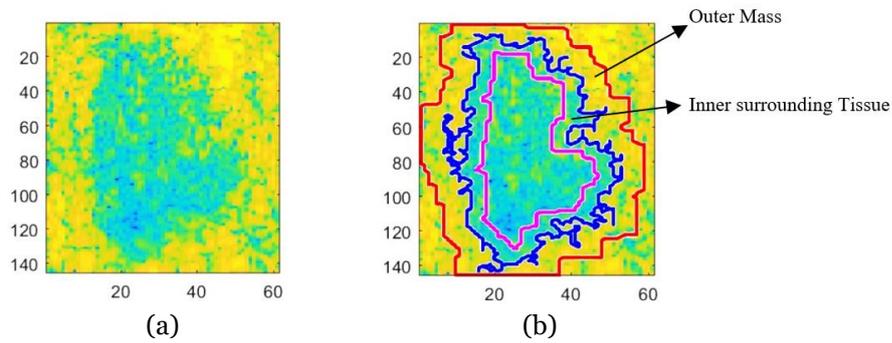

Fig. 11. Example of automatically segmentation of the lesion boundaries for inner and outer mass at the width of 6 pixels respectively. (a) Breast US image (b) Three boundaries considering inner and outer mass denoted by pink and red color boundaries respectively where the blue one is the lesion boundary.

## Tilted Ellipse Area ($A_{ellipse}$)

The area bounded by an ellipse is expressed as:

$$A_r = \pi ab \tag{27}$$

Where, 'a' and 'b' are the horizontal and vertical sub axis lengths inside the best fit tilted ellipse respectively as shown in Fig. 12.

## Tilted Ellipse Compactness ($CP_{ellipse}$)

This feature is estimated using the ration of *tilted ellipse perimeter* and the *tilted ellipse area* to determine how close the pixels of lesion region are to the center of lesion region.

So, the compactness is measured as,

$$CP_{ellipse} = \frac{P_{ellipse}^2}{A_{ellipse}} - 1.0 \qquad (28)$$

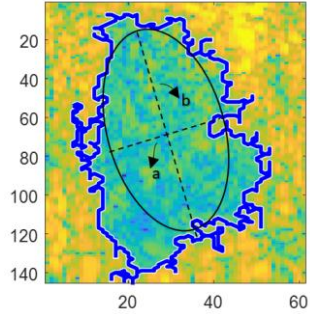

Fig. 12. The sub axis denoted by 'a' and 'b' inside the best-fit tilted ellipse.

*Power spectral density (PSD) Peaks ($P_{PSD}$)*

The Yule-Walker method is applied separately on the pixel matrix of the vertical sub axis 'b' and horizontal sub axis 'a' to determine the Power Spectral Density (PSD). The values at which the peaks in the PSD occur are observed [43]. The first three peak values are extracted for classification purposes. An example of simulated PSD spectrum of a WCP image at contourlet decomposition level P4D16 is shown in Fig. 13.

*Kolmogorov-Smirnov statistics (klv)*

The performance of the RiIG *pdf* is compared with that of Nakagami *pdf* using the well-known Kolmogorov-Smirnov (KS) statistics (*klv*) and variance stabilized percentile probability plot (*pp-plot*) as shown in Fig. 5. The KS statistics is given by

$$klv = \max_{x \in R} |F_e(x) - F_a(x)| \qquad (29)$$

where, *klv*, $F_a(x)$ and $F_e(x)$ denote the Kolmogorov-Smirnov (KS) statistic values, cumulative density function (*cdf*) of the prior three *pdf*s and the empirical cumulative density function (*cdf*) respectively [28].

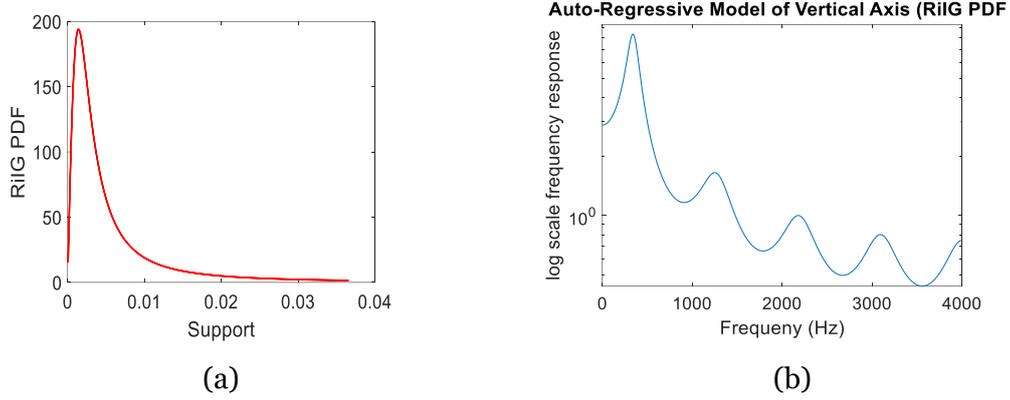

Fig. 13. Example of power spectral density (PSD) peaks, simulated at contourlet decomposition level P4D16 for a vertical sub-axis of tilted ellipse considering WCP image of a malignant mass (a) RiIG *pdf* (b) PSD peaks.

*Percentile Probability Plot (pp-plot)*

The *pp-plot* is simulated (Fig. 5) by plotting $F_a(x)^t$ against $F_e(x)^t$ where a linear plot means excellent fitting [29].

$$F_a(x)^t = \frac{2}{\pi} arcsin\left(\sqrt{F_a(x)}\right) \quad (30)$$

$$F_e(x)^t = \frac{2}{\pi} arcsin\left(\sqrt{F_e(x)}\right) \quad (31)$$

Here, *t* denotes the transpose operation.

*Kullback–Leibler divergence (klb)*

The suitability of the RiIG *pdf* is also compared with the Nakagami *pdf* using the well-known Kullback–Leibler (KL) divergence. The KL divergence is given by [35]

$$KL(P_{emp}, P) = \int P_{emp}(x) \log_2 \frac{P_{emp}(x)}{P(x)} dx \quad (32)$$

Here, $P(x)$ denote the probability density function (*pdf*) of the Nakagami and RiIG statistical models and $P_{emp}(x)$ denote the empirical *pdf*.